\documentstyle{article}
\begin{document}
{\large
\begin{center}
{\Large \bf ON THE KINEMATICS OF A COROTATING
RELATIVISTIC PLASMA STREAM IN THE PERPENDICULAR
ROTATOR MODEL OF \\ A PULSAR MAGNETOSPHERE}
\vskip 27pt
{\large  O.V.~Chedia, T.A.~Kahniashvili, G.Z.~Machabeli \& \\
\vskip 7pt
I.~S.~Nanobashvili}
\vskip 7pt
{\it Department of Theoretical Astrophysics, Abastumani Astrophysical\\
Observatory, A.Kazbegi ave.$2^a$, 380060 Tbilisi, Republic of Georgia\\}
\end{center}
\vskip 327pt
\newpage
\baselineskip=18pt
\begin{center}
{\bf Abstract}
\end{center}

An investigation of the kinematics of a rotating relativistic plasma stream
in the perpendicular rotator model of the pulsar magnetosphere is presented.
It is assumed that the plasma (ejected from the pulsar) moves along the
pulsar magnetic field lines and also corotates with them. The field lines
are considered to be radial straight lines, located in the plane which is
perpendicular to the pulsar rotation axis. The necessity of taking particle
inertia into account is discussed. It is argued that the "massless"
("force-free") approximation cannot be used for the description of this
problem. The frame selection is discussed and it is shown that it is
convenient to discuss the problem in the noninertial frame of ZAMOs (Zero
Angular Momentum Observers). The equation of motion and the exact set of
equations describing the behaviour of a relativistic plasma stream in the
pulsar magnetosphere is obtained. The possible relevance of this investigation
for the understanding of the formation process of a pulsar magnetosphere is
discussed.
\newpage
\begin{center}
{\bf Introduction}
\end{center}

In connection with several astrophysical problems, e.g. the theoretical
description of pulsar magnetospheres, the question of investigation of plasma
ejection from a rotating source (e.g. a star) along its magnetic field
lines is arising. It is most important to investigate the behaviour of
this stream  near the pulsar surface.

While investigating the pulsar wind, the change of particle momentum is
usually~~neglected.~~The~~equations~~are~~then~~formulated~~in~~the~~"massless" \\
approximation ${{m}{=}{0}}$ (Mestel, 1973; Endean, 1974), in which
the particles are moving along the pulsar magnetic field lines like
photons moving along geodetic lines. It seems to us indubitable, that the
inertia of the particles, disregarded in a large number of previous papers,
must have a great influence on the pulsar magnetosphere formation process.
The importance of taking into consideration the inertia of the particles
was understood a long time ago and treated as a perturbation to "massless"
approximation (Scharlemann, 1974; Henriksen and Norton, 1975). It seems to
us that giving up the "massless" approximation not only shows the essence of
the problem more adequately, but also simplifies the problem from the
mathematical point of view.

We use the perpendicular rotator model of the pulsar magnetosphere and
treat only the polar cap. It is assumed that the pulsar magnetic field
lines are the radial straight lines. In our paper is discussed the case, when
the plasma particles corotate with the pulsar magnetic field lines.
We also assume that the ${E_{\Vert}}=0$, i.e. the electric field component
along the magnetic field lines is zero.

We will choose the simplified geometry, discuss
the frame selection and the derivation of the equation of
motion of a relativistic stream. We will argue that the "massless"
("force-free") approximation cannot be used for the description of this
problem and discuss the difference of our statement of the problem from those
discussed before by a number of authors (Mestel, 1973; Endean, 1974;
Henriksen and Norton, 1975; Michel, 1969; Kennel et al., 1983 and others).

We will obtain the equation of motion of a relativistic plasma stream and
the exact set of equations for the description of the behaviour of the
stream.

\begin{center}
{\bf Basic Considerations}
\end{center}

Let us consider the simplified geometrical model, when the pulsar rotation axis
is perpendicular to its magnetic momentum. The magnetic field lines are
considered as radial straight lines, located in the plane which is
perpendicular to the pulsar rotation axis. If we discuss the processes in the
region which has the linear size ${{d}{\ll}{R_c}}$, where ${R_c}$ is the
curvature radius of the magnetic field lines, this approximation is justified.
The ejected particles not only move along the radius ${r}$, but also coroatate
with the pulsar magnetosphere because the field lines are corotating with the
plasma:

$$
{{{\vec {E}}+
{[{\vec {V}}{\vec {B}}]}}{=}{0}}. \eqno (1)
$$

Here and below we use geometric units ${{c}{=}{G}{=}{1}}$.

The condition (1) means, that because of the rotation of the magnetic field
${\vec {B}}$, a field ${\vec {E}}$ is generated, which forces plasma
particles to corotate with the magnetic field lines with linear velocity:

$$
{{\vec {V}}{=}{[{\vec {\Omega}}{\vec {r}}]}}, \eqno (2)
$$
where ${\vec {\Omega}}$ is the angular velocity vector of the star.

In the "massless" approximation, in an inertial frame there are no other
forces in the equation of motion than the Lorentz force ${{e}{({\vec {E}}+
{[{\vec {V}}{\vec {B}}]})}}$. The centrifugal force is neglected. However,
by condition (1) the Lorentz force vanishes and ${{{{d}{\vec {p}}}\over
{{d}{t}}}{=}{0}}$. According to Rylov (1989) the
equation ${{{{d}{\vec {p}}}\over {{d}{t}}}{=}{0}}$ in
approxi\-mation ${{m}{=}{0}}$ can
be justified in two cases: ${{\vec p}{\ne}{0}}$ and ${{\gamma}{\rightarrow}
{\infty}}$, or finite ${\gamma}$ and ${{\vec p}{=}{0}}$. The first case
is called by Rylov "force-free" app\-roximation and the se\-cond
one --- "massless" approximation. But to disregard the centrifugal
force it must be shown under which conditions this is a valid approximation.
Below we will show, that disregarding the centrifugal force is equivalent to
giving up corotation. In order to understand the physical reason for
corotation, let us consider a well known example: let us make a conductor
move across a magnetic field with a constant velocity ${\vec {V}}$.
Then, the Lorentz force ${{e}{[{\vec {V}}{\vec {B}}]}}$ will act in
dif\-fe\-rent directions on the charges of different signs of the conductor
which will cause a charge separation, i.e. generation of the electrostatic
field. Due to the separation of charges, some energy of the source
making the conductor move across a magnetic field lines with constant
velocity (this velocity satisfies the condition (1)), must be expended.
An analogous situation holds in the case
when an observer moves with respect to the conductor with the velocity
${\vec {V}}$. The condition (1) allows also such an interpretation: the motion
of the particles (of any sign) perpendicularly to the magnetic field is
possible if an electric field ${\vec {E}}$ ${({{\vec E}{\bot}{\vec
B}})}$ exists and the velocity of this motion is equal to the electric
drift velocity ${{\vec {V}}{=}
{{[{\vec {E}}{\vec {B}}]}\over {B^2}}}$.

Now let us assume that the outer source makes the conductor move with an
acceleration. Then it is evident, that the work of the source is expended not
only on the charge separation, but also on the acceleration of the conductor.
Therefore, except of the Lorentz force, there will be an additional force
${\vec {F}}$ in the equation of motion, but due to the condition (1) it
will take the following form:

$$
{{{{d}{\vec {p}}}\over {{d}{t}}}{=}{\vec {F}}}.
$$

Returning to our problem we conclude: the pulsar rotation makes the magnetic
field lines rotate, this rotation generates the electric field, which is
oriented perpendicularly to the magnetic field
lines and because of the condition
(1) makes the charged particles (ejected from the star) corotate with the
magnetic field lines. During this motion,
the particle velocity changes its direction,
therefore a force is acting on the particles. This force in our case is
directed along the magnetic field lines and is not compensated.

Considering analogous problems in classical mechanics, for example the
motion of a bead located in a rotating tube, the reaction force
appears in the equation of motion. This force makes
the bead not only rotate together with the tube, but also accelerates it along
the tube.

In our problem, according to the condition (1)
the magnetic field lines and the generated electric field
act like the rotating tube for the charged
par\-ticles. It is more convenient to find the form of the reaction force in a
noninertial frame, which rotates together with the particles. The situation
described above is very similar to the problems connected with the
description of particle motion in curved space-time, in particular in the
gravitational field of the black hole --- in the Schwarzschild or Kerr metric,
because according to the Einstein principle of equivalence, locally we cannot
distinguish gravitation from non\-inertiality. Using the principle of
equivalence, the (3+1) formalism is suitable for our problem. The (3+1)
formalism is described in Thorne et al. (1988).
In this formalism instead of four-vectors, three spatial and one
time coordinates are used.

The problem of the motion of particles in the pulsar magnetosphere can be
solved in the local-inertial frame of the observers, who are measuring the
physical quantities in the immediate vicinity of themselves. They are called
the Zero Angular Momentum Observers (ZAMOs). ZAMOs cover the whole space-time.
Each of the observers uses its proper time ${\tau}$, which is different from
the universal time ${t}$. We must mention that the proper time of the
observer who moves together with the particle, is different from the
proper time of ZAMOs ${\tau}$.

The transition from the inertial frame to the frame, which is connected with
the pulsar magnetosphere rotation can be done by following relation:

$$
{{{t}{=}{t'}}, {{\varphi}{=}{{\Omega}{t}}}, {{r}{=}{r'}}, {{z}{=}{0}}}.
\eqno (3)
$$

The metric, describing such a frame has the following form:

$$
{{{d}{S^2}}{=}{{-}{(1-{{{\Omega}^2}{r^2}})}{{d}{t^2}}{+}{{d}{r^2}}}}.
\eqno (4)
$$

>From (4) it follows, that the interval of the proper time of ZAMOs is
connected with the universal time interval as

$$
{{{d}{\tau}}{=}{\alpha}{{d}{t}}}, \eqno (5)
$$
where ${{\alpha}{=}{\sqrt {1-{{{\Omega}^2}{r^2}}}}}$ and it is called the
"lapse function".

A photon, propagating from the periphery to the center of rotation, from
observer to observer would be "reddening". The reason is that the rate
of the time lapse for each of the ZAMOs differs from those for the
others and also differs from the lapse of the universal (laboratory)
time. The lapse function ${\alpha}$ not only connects the proper time
of ZAMOs ${\tau}$ with the universal time ${t}$,
but also expresses the gravitational potential, i.e. the inertial force.
In the frame of ZAMOs the particle has the gravitational acceleration

$$
{\vec g}{=}{-}{{{\vec {\nabla}}{\alpha}} \over {\alpha}}. \eqno (6)
$$

The equation of motion of the particle in the frame of ZAMOs takes the
following form:

$$
{{{d}{\vec p}}\over {{d}{\tau}}}{=}{\gamma}{\vec g}{+}
{{e}\over {m}}{({\vec E}{+}{[{\vec V}{\vec B}]})}, \eqno (7)
$$
where ${\gamma}{=}{{1}\over {\sqrt {{1}{-}{V^2}}}}$ is the Lorentz-factor and
${\vec V}{=}{{1}\over {\alpha}}{{{d}{\vec r}}\over {{d}{t}}}$ is the
velocity of the par\-ticle, ${\vec p}$ is
the dimensionless momentum, ${\vec p}{\rightarrow}{{\vec p}\over m}$.

We must mention, that the equation of motion contains the acceleration
${\vec g}$ from
(6) only in the frame, for which
${d}{{\varphi}'}{=}{0}$. In the generalized rotating frame, where
${d}{{\varphi}'}{\ne}{0}$, according to the (3+1) formalism, the lapse
function ${\alpha}{=}{1}$, but instead of
the acceleration (6) the "gravimagnetic" acceleration appears
(Thorne et al. 1988), which must be
written in the equation of motion. This equation differs from (7) only by the
redesignation of the angular velocities and, of course, describes the same
process and gives the same results as (7).

While considering the hydrodynamical problem, according to the (3+1)
formalism, instead of the equation of motion for a single particle, the
energy-momentum tensor conservation (the law of force balance) is used.
Also, using the particle consertvation law (the continuity equation) and
neglecting the hydrodynamical stream pressure, the equation, which describes
the stream motion takes the following form:

$$
{{{1}\over {\alpha}}{{{\partial}{\vec p}}\over{{\partial}{t}}}{+}
{({\vec V}{\vec {\nabla}})}{\vec p}{=}{-}{\gamma}
{{{\vec {\nabla}}{\alpha}} \over {\alpha}}{+}
{{e}\over{m}}{({{\vec {E}}+
{[{\vec {V}}{\vec {B}}]}})}}, \eqno (8)
$$

In the equation (8) the derivative ${{d}\over{{d}{\tau}}}$ is changed
with the convectional derivative
${{d}\over{{d}{\tau}}}{\rightarrow}{{1}\over {\alpha}}
{{\partial}\over{{\partial}{t}}}{+}{({{\vec V}{\vec {\nabla}}})}$
and the term proportional to the pressure gradient
${{\vec {\nabla}}{P}}$ (${P}$ is the pressure) is neglected.

Let us discuss the difference between equation (8) and the
equ\-ation, usually used during the consideration of similar
problems in the magnetohydrodyna\-mical approximation. In previous papers, the
ac\-celeration ${{{\vec {\nabla}}{\alpha}} \over {\alpha}}$ was
disregarded. The problem was formulated in the inertial frame.
It is easy to show, that the momentum ${\vec p}$ in the
frame of ZAMOs coincides
with the momentum ${{\vec p}~'}$ in inertial frame. Really, according to
the definition
${\vec p}{=}{\gamma}{\vec V}$, where
${\vec V}{=}{{1}\over {\alpha}}{{{d}{\vec r}}\over {{d}{t}}}$,
and

$$
{{\gamma}{=}{\alpha}{{\gamma}~'}}, \eqno (9)
$$
we will find, that ${{\vec p}~'}{=}{{\vec p}}$, because
${{\vec V}'}{=}{{{d}{{\vec r}~'}}\over {{d}{t}}}{=}{{{d}{\vec r}}\over
{{d}{t}}}$.
It is easy then to rewrite the
equation (8) for the quantities defined in inertial frame. Omitting the prime
for all quantities, we will obtain

$$
{{{{\partial}{\vec p}}\over {{\partial}{t}}}{+}
{({\vec V}{\vec {\nabla}})}{\vec p}{=}{-}{\gamma}{\alpha}
{\vec {\nabla}}{\alpha}{+}{{e}\over {m}}{({{\vec {E}}+
{[{\vec {V}}{\vec {B}}]}})}}. \eqno (10)
$$

The equation (10) contains the force ${\vec F}{=}{-}{\gamma}{\alpha}
{\vec {\nabla}}{\alpha}$,
which was disregarded in previous
papers. It is the analogy of the force, which causes the centrifugal effect
and is equal to

$$
{{\vec F}{=}{{{{\Omega}^2}{\vec r}}\over
{\sqrt
{{1}{-}{{{\Omega}^2}{r^2}}{-}{{({{{d}{r}}\over{{d}{t}}})}^2}}}}}.
\eqno (11)
$$
This force exceeds ${\vec {\nabla}}{P}$ (${P}$ is the pressure), which was
taken into
consideration in the paper by Kennel et al. (1983).
To find the force ${\vec F}$,
it is not necessary to write the equation in the
noninertial frame. Staying in the
inertial frame
and taking into consideration that the axially symmetric magnetic field
changes its direction in the space, i.e.
${\vec B}{=}{\vec B}{({\vec r})}{=}{B_r}{\left ( {{\vec r}\over
{\mid {\vec r} \mid}} \right)}$, the equation of motion for the
guiding center can be obtained. The relativistic generalization of this
problem (which was stated and solved for the first time by Alfven) can be
found in Sivukhin (1963).
We must mention, that in the paper of Cohen and Rosenblum (1972)
the problem was
investigated in the rotating frame and the equations were written for the
four-vectors, but then the exact equations for the motion of the particles
were not used.

Besides, there are two principal differences of our consideration from
previous ones (e.g. Michel, 1969; Kennel et al., 1983).
In the problem discussed here the extremely large surface magnetic field
(${B_0}{\simeq}{{10}^{12}}{-}{{10}^{13}}$ G) of the neutron star is
essential. On the other hand, in the magnetosphere outside the star
a magnetic field ${\vec {B_1}}$ can be generated.
The source of the
field ${\vec {B_0}}$ is located inside the star, but the source of the field
${\vec {B_1}}$ is a
current ${\vec j}$, which is generated in the
magnetosphere. The current
${\vec j}$ has no influence on the source of the pulsar magnetic field
${\vec {B_0}}$ and we will
consider ${\vec {B_0}}$ as an external magnetic field, in which the
plasma is located.

Furthermore, in the previous papers another assumption was
made, namely that the inertia of the particles, i.e. the
term which contains the partial time derivative
${{\partial}\over {{\partial}{t}}}$, may be disregarded. In
Henriksen and Norton (1975) the
derivative ${{\partial}\over {{\partial}{t}}}$
for the transversal components of the quantities was changed
to the derivative with respect to the azimuthal coordinate,
${{\partial}\over {{\partial}{t}}}
{\rightarrow}{\Omega}
{{\partial}\over {{\partial}{\varphi}}}$ and the change of the
quantities along the magnetic field was disregarded.

The quantities in
equation (10) can be written as:

$$
{{{\vec E}{=}{\vec {E_0}}{+}{\vec {E_1}}}, \hskip 1cm
{{\vec B}{=}{\vec {B_0}}{+}{\vec {B_1}}}, \hskip 1cm
{{\vec p}{=}{\vec {p_0}}{+}{\vec {p_1}}}}, \eqno (12)
$$
where ${\vec {E_0}}$, ${\vec {B_0}}$ and ${\vec {p_0}}$ are the basic
terms, and ${\vec {E_1}}$, ${\vec {B_1}}$ and ${\vec {p_1}}$ are the
perturbations in the first approximation of expansion over the parameter
of weak turbulence (${{{E_1}^2}\over {{m}{n}{\gamma}}}$).

>From the condition (1) in zeroth approximation we have:

$$
{{{\vec {E_0}}+
{[{\vec {V_0}}{\vec {B_0}}]}}{=}{0}}
$$
and from the equation (10) for the radial acceleration we will obtain
(Machabeli and Rogava, 1994):

$$
{{{d^{2}r}\over{dt^2}}={{{\Omega}^{2}r}\over
{1-{\Omega}^{2}r^2}}{\left[1-{\Omega}^{2}r^2-
2{\left({{dr}\over
{dt}}\right)}^2 \right]}}. \eqno (13)
$$

Here we discuss only the radial motion, because the azimuthal one is
determined by corotation $({V_{\varphi}}={\Omega}{r})$.
The solution of the equation (13)
can be presented in the form:

$$
{{r}{({t})}{=}{{V_0}\over {\Omega}}{{{Sn}{{{\Omega}{t}}}}\over
{{dn}{{{\Omega}{t}}}}}}, \eqno (14)
$$
where ${Sn}$ and ${dn}$ are the Jacobian elliptical sine and modulus
respectively (Abramovitz and Stegun, 1964),
${V_0}$ is the initial velocity of the particle. From the equation (13)
it follows
that if the radial velocity ${{V_r}>{{1}\over {\sqrt {2}}}}$,
the acceleration changes its sign and the
particle is not accelerated, but braked (see in
detail Machabeli and Rogava (1994)).

Using the asymptotic expression for the Jacobian function we find that
when ${V_0}{\rightarrow}{1}$,

$$
{{r}{({t})}{=}{{V_0}\over {\Omega}}{sin}{\Omega}{t}}. \eqno (15)
$$
For the radial velocity
we obtain:

$$
{{{V_0}_r}{=}{V_0}{cos}{\Omega}{t}}, \eqno (16)
$$
from which it follows that

$$
{{{V_0}^2}{=}{{({{V_0}_r})}^2}{+}{{({{V_0}_{\varphi}})}^2}{=}
{const}}. \eqno (17)
$$

This fact is not quite evident. In the nonrelativistic case, as it is known,
the particle kinetic energy increases exponentially at the expense of the
external source, which keeps constancy of rotation. This follows from the
general solution (14) too, when ${V_r}{\ll}{1}$;
expanding the Jacobian functions we
obtain:

$$
{{r}{({t})}{=}{{V_0}\over {\Omega}}{sh}{\Omega}{t}}, \eqno (18)
$$
where ${sh}$ is the hyperbolic sine.

The principle difference between nonrelativistic and relativistic motion can
be easily understood if we rewrite the equation (10) in a
form different from (13):

$$
{{{{\gamma}_0}{{{d}{{V_0}_r}}\over {{d}{t}}}{+}
{({{\gamma}_0}{{\Omega}^2}{r}{+}{{\gamma}_0}
{{{d}{{V_0}_r}}\over {{d}{t}}})}
{{{\gamma}_0}^2}{{{V_0}_r}^2}{=}{{\gamma}_0}{{\Omega}^2}{r}}}. \eqno (19)
$$

The second and third terms in the left side describe the change of mass
(${\gamma}$-factor) in time. It is evident (from (19)) that for
small velocities
${{V_0}_r}{\rightarrow}{0}$
(${{\gamma}_0}{\rightarrow}{1}$) we will obtain
${{{d}{{V_0}_r}}\over {{d}{t}}}{-}{{\Omega}^2}{r}{=}{0}$ and the
solution (18).
But if ${{V_0}_r}{\rightarrow}{1}$
(${{\gamma}_0}{\rightarrow}{\infty}$),
we will obtain ${{{d}{{V_0}_r}}\over {{d}{t}}}{+}{{\Omega}^2}{r}{=}{0}$
and the solution (16). From the equation (19) also follows
that the term connected with the change of mass dominates over the
other terms if

$$
{{{V_0}_r}>{\sqrt {{{1}{-}{{{V_0}_{\varphi}}^2}}\over {2}}}}. \eqno (20)
$$

When (20) is valid, the inertia of particles is such,
that the centrifugal force changes its sign and the motion is braked.
We can see, that when ${{V_0}_r}{\rightarrow}{1}$,
the "massless" ("force-free") approximation can be justified  only if
"force-free" means ${{{d}{{{\gamma}_0}}}\over {{d}{t}}}{=}{0}$.
If we disregard the right side in the
equation (19) in comparison with the first term of the left side, this fact
automatically means disregarding the first term in comparison with the
second term in brackets on the left side of the equation.
Then ${{p_0}_r}{=}{const}$ --- the particle motion is uniform and
straightforward and we have no corotation. Therefore disregar\-ding the
force ${\vec F}$ is possible only in the case ${\Omega}{=}{0}$.

Thus the "massless" ("force-free") approximation cannot be used for the
"solid-body" type rotating ${({\vec {V}}{=}{[{\vec {\Omega}}{\vec
{r}}]})}$ relativistic plasma stream.
${{{d}{{p_0}_r}}\over {{d}{t}}}{\ne}{0}$, especially
${{{d}{{p_1}_r}}\over {{d}{t}}}$ must
not be equal to zero.

If we add to the equation (10) the continuity equation and Maxwell
equations we shall have a set of equations, which completely
describes the behaviour of a relativistic plasma stream corotating
with the pulsar magnetosphere:

$$
{{{{\partial}{\vec p}}\over {{\partial}{t}}}{+}
{({\vec V}{\vec {\nabla}})}{\vec p}{=}{-}{\gamma}{\alpha}
{\vec {\nabla}}{\alpha}{+}{{e}\over {m}}{({{\vec {E}}+
{[{\vec {V}}{\vec {B}}]}})}}
$$

$$
{{{\partial{\rho}}\over {\partial t}}=
-{{\vec{\nabla}}{({{\alpha}{\vec j}})}}}
$$

$$
{({{\vec {\nabla}}{\vec B}})}=0
$$

$$
{({{\vec {\nabla}}{\vec E}})=4{\pi}{\rho}}
$$

$$
{{{\partial {\vec B}}\over {\partial t}}=
-{[{\vec {\nabla}}{({\alpha}{\vec E})}]}}
$$

$$
{{{\partial{\vec E}}\over {\partial t}}=
{{[{{\vec \nabla}}{({\alpha}{\vec B})}]}-{4{\pi}{\alpha}{\vec j}}}}
$$

where ${\rho}$ and ${\vec j}$ are the charge and current densities
respectively.

\begin{center}
{\bf Conclusion}
\end{center}

We have investigated the role of particle inertia and shown that the
"massless" ("force-free") approximation cannot be used for the "solid-body"
type rotating ${({{\vec {V}}{=}{[{\vec {\Omega}}{\vec {r}}]}})}$ relativistic
plasma stream description. We must mention that our investigation was
carried out under the assumption ${E_{\Vert}}=0$ and if ${E_{\Vert}}{\ne}0$,
this would probably have the strong influence on our results. We also derived
the equation of motion and a set of equations for the description of a
relativistic plasma stream in the perpendicular rotator model of pulsar
magnetospheres.

\begin{center}
{\bf Acknowledgements}
\end{center}

We would like to thank the anonymous referee for constructive
suggestions and criticism.

\newpage
\begin{center}
{\large \bf References}
\end{center}

\begin{flushleft}
\begin{tabular}{ll}
& Abramowitz M. and Stegun I.A.: "Handbook of Mathematical Functions", 1964. \\
& Cohen J. and Rosenblum A.: 1972, Astrophys. Space Sci., {\bf 6},130. \\
& Endean V.G.: 1974, Astrophys. J., {\bf 187},359. \\
& Henriksen R.N. and Norton J.A.: 1975, Astrophys. J., {\bf 201},719.\\
& Kennel C.F., Fujimura F.S. and Okamoto I.: 1983, Geophys. Astrophys.Fluid \\
& Dynamics, {\bf 26},147.\\
& Machabeli G.Z. and Rogava A.D.: 1994, Phys.Rev., {\bf 50},98.  \\
& Mestel L.: 1973, Astrophys. Space Sci., {\bf 24},289.\\
& Michel F.C.: 1969, Astrophys. J., {\bf 153},717. \\
& Rylov Yu.A.: 1989, Astrophys. Space Sci., {\bf 158},297. \\
& Scharlemann E.T.: 1974, Astrophys. J., {\bf 193},217.\\
& Sivukhin D.V., "Dreifovaia Teoria dvijenia zariajennoi chasticy v elektromagnit-\\
& nykh Poliakh"; "Voprosy Teorii Plazmy" (pod redakciei Leontovicha M.A.), \\
& tom 1, Atomizdat, M., 1963. (in russian) \\
& Thorne K.S., Price R.H., Macdonald D.A. "Black Holes: The Membrane \\
& Paradigm" (Yale University Press, New Haven and London) 1988.\\
\end{tabular}
\end{flushleft}

\end{document}